\documentclass[floatfix,showpacs,twocolumn,pre]{revtex4}
\usepackage{pstricks,pst-node,pst-text,pst-3d}
\usepackage{amsmath,amsfonts,amssymb,graphicx}

\providecommand{\U}[1]{\protect\rule{.1in}{.1in}}
\newtheorem{theorem}{Theorem}
\newtheorem{acknowledgement}[theorem]{Acknowledgment}

\begin{document}

\preprint{}
\title[Optimization and Scale-freeness]{Optimization and Scale-freeness for
Complex Networks}
\author{Petter Minnhagen}
\author{Sebastian Bernhardsson}
\affiliation{Dept.of Physics, Ume\aa\ University, 901 87 Ume\aa , Sweden, \\
Center for Models of Life, Copenhagen, Denmark}
\pacs{PACS number}

\begin{abstract}
Complex networks are mapped to a model of boxes and balls where the balls
are distinguishable. It is shown that the scale-free size distribution of
boxes maximizes the information associated with the boxes \emph{provided}
configurations including boxes containing a finite fraction of the total
amount of balls are excluded. It is conjectured that for a connected network
with only links between different nodes, the nodes with a finite fraction of
links are effectively suppressed. It is hence suggested that for such
networks the scale-free node-size distribution maximizes the information
encoded on the nodes. The noise associated with the size distributions is
also obtained from a maximum entropy principle. Finally explicit predictions
from our least bias approach are found to be born out by metabolic networks.
\end{abstract}

\maketitle

\textbf{Networks galore! Representations of real complex systems in terms of
networks range over all science from social science, economics, internet,
physics, chemistry and biology and much more. Basically a network is a
representation of who or what is connected to or influences whom or what.
It takes the form of some irregular cobweb where the parts (the who's or
what's) are connected by links. The parts are called nodes so the
representation is in terms of nodes and links (see Fig.1). One feature of a
network is the number of links which are attached to a node: a network can
be characterized by $N(k)$, the number nodes with $k$ links attached to
them. In many real networks one finds that this distribution over sizes $k$
is very broad and power law like i.e. $N(k)\sim k^{-\gamma }$. Why is this
and what does it imply? This is still to large extent an open question. Here
we adress this question using the maximum entropy principle. The predictive
value of this principle is greatest when it fails! For example, when it was
found that the measured blackbody radiation did have a smaller entropy than
the one predicted from Maxwells equations and classical statistical
mechanics, the maximum entropy principle (had it been known at the time)
would immediately tell you that a physical constraint was lacking in the
theory. This alas turned out to be the Planck constant. We use the same
reasoning here: we first find that the maximum entropy for an unconstrained
network does have a larger entropy than the broad distribution found in real
systems. So there is a missing contraint! We argue that this missing
contraint is the advantage (in many cases) of maximizing the information
encoded on the nodes. Thus we are suggesting that the real advantage is not
to maximize the total number of possibilities but rather to maximize the
information encoded on the nodes. Our "least bias" approach gives explicit
predictions for real networks which can be tested. We demonstrate that
metabolic networks (a network representation of the metabolism in a cell)
are likely to be a maximum information network.}

Complex networks have undergone a rapid surge of interest and a number of
review articles already exist \cite{albert02}\cite{dorog03}\cite{newman03}%
\cite{bocca06}. A striking observation in this field is that many real
networks have a broad scale-free like degree distribution. Why is this and
what does it imply for the evolution mechanism of the networks? There exists
many suggestions \cite{albert02}\cite{dorog03}\cite{newman03}\cite{bocca06}.
Most suggestions concerns steadily growing networks. The most well-known
proposal for an evolution mechanism which produces a scale-free network for
this case is the preferential attachment scenario \cite{barabasi99}. This
mechanism rests on two explicit assumptions: 1) a link on a new node
attaches randomly to a node belonging to the network with a probability
directly proportion to the number of links that are already attached to this
latter node. 2) once a link is in place it stays in place. Although this
model mechanism has become a successful prototype explanation in many cases 
\cite{albert02}, its applicability is obviously limited by the very specific
model assumptions. Other suggestions are based on various optimization ideas 
\cite{valverde02}, and, finally, some concern non-growing networks 
\cite{beom05}\cite{thurner05}\cite{Bianconi}.

In the present paper we discuss non-growing networks. Such a network has a
fixed number of nodes and links and the time-evolution is through rewiring
of links. First we discuss the definition of a random network. Next we
address the question "What is the least bias needed to impose on a random
network in order to obtain a scale-free distribution?". The possible "least
bias" solutions are presented together with the corresponding noise. Finally
characteristic features arising from the "least bias" random scale-free
network are extracted and compared to real metabolic networks. Some
speculations are added. Our approach is based on information theory and
statistical mechanics \cite{jaynes57}.

\section{Randomness and States}

\begin{figure}[tbp]
\begin{center}
\includegraphics[width=\columnwidth]{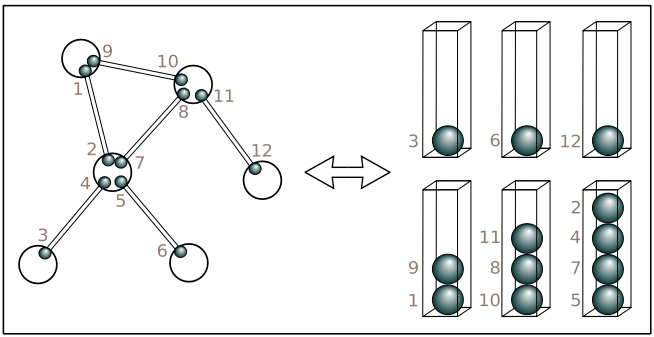}
\end{center}
\caption{Figure showing how an undirected network can be described by
numbered balls placed in boxes. The balls represent link-ends and are
distinguishable because each link start and end on a specific node.}
\label{box}
\end{figure}

In statistical mechanics random means that each possible state of a system
is equally probable. Consequently, one first has to identify what is meant
by a state before the concept of randomness takes on a precise meaning. Here
we will follow Ref.\ \cite{jaynes57} by making the connection between states
and different distinguishable ways of distributing objects. For example the
lower right hand box in Fig.1 contains four distinguishable (numbered)
balls. There are $4! = 24$ ways in which these four balls can be distributed
into the box. All these ways are distinguishable because the balls are
distinguishable and thus correspond to $24$ different states. This is just
like the box contained four different compartments each of which can contain
one ball.

Figure 1 shows a non-directed network in terms of undirected links and
nodes. Each link starts and ends with a link-end. Let us consider the
link-ends on a node. In statistical mechanics one could think of the
link-ends as particles. Since each link-end on a particular node is
connected to another link-end on a different node, it means that the
link-ends on a node are de facto distinguishable. Thus the closest analogy
with statistical mechanics is that of boxes (nodes) containing
distinguishable balls (link-ends). A first description of the system is then
in terms of balls in boxes \cite{bialas00}\cite{burda01}. In addition there
are constraints imposed by the network topology. Typical constraints are no
loops (links starting and ending on the same node), at most a single link
between two nodes, and keeping the network connected. A necessary (but not
sufficient) constraint corresponding to connected networks is that each box
always contains at least one ball. The other two constraints are often
unimportant and occurs with a vanishing probability in the limit of large
random networks \cite{dorog03}. However, sometimes the no-loop constraint is
crucial, as will be discussed further on.

\section{Balls in Boxes}

Let us start with randomly distributing $M$ \textit{distinguishable} balls
into $N$ boxes. The total number of ways of picking the balls are $M!$ and
the outcome is a distribution into the $N$ boxes where $N(k)$ boxes has $k$
balls. Since all the balls are distinguishable all boxes with $k$ balls are
distinguishable and consequently the number of distinguishable ways of
distributing the balls into the boxes, $\Omega ,$ are given by

\begin{eqnarray}
\ln \Omega &=&\ln M!-\sum_{k=0}\ln N(k)! \\
&\eqsim &M\ln M-M-\sum_{k=0}N(k)[\ln N(k)-1]
\end{eqnarray}

The degeneracy expressed by $N(k)!$ reflects the trivial fact that
two boxes which contain the same number of balls have indistinguishable
sizes. \textit{The connection to statistical physics is the one-to-one
correspondence between distinguishable ways of distributing the balls and
the states of the system} \cite{jaynes57}. The unbiased estimate of the
total number of different states $\Omega $ is given by the maximum of $\ln
\Omega $ with respect to the distribution $N(k),$ subject to the constraints 
$\sum_{k=1}^{N}N(k)=N$ and $\sum_{k=1}^{N}kN(k)=M$. This gives by
variational calculus $-\ln N(k)+a-bk=0$ with the solution 
\begin{equation}
N(k)=A\exp (-bk)
\end{equation}%
[$a$ and $b$ are the Lagrange multipliers corresponding to the constraints
and $A$ is a normalization constant corresponding to $\sum_{k}N(k)=N$]. This
is the exponential distribution which within statistical physics in other
variables can be recognized as the well-known Boltzmann distribution.

Suppose now that the balls instead are \textit{indistinguishable}. In this
case the boxes can only be distinguished if the number of balls they contain
are different. Compared to the case with distinguishable particles this
means that all ways of distributing $k$ balls into a box are
indistinguishable or degenerate. We define the degeneracy $f(k)$ as the
number of of ways of distributing balls into a box which are
indistinguishable. This means that $f(k)$ is the total number of of ways of
distributing $k$ different particles into a box divided by the total number
of distinguishable states which the box can have. In the present example
this is obviously $f(k)=k!/1$. Consequently total number of distingushable
state are now

\begin{align}
\ln \Omega =& \ln M!-\sum_{k=0}\ln N(k)!-\sum_{k=0}N(k)\ln f(k)]  \notag \\
\eqsim & M\ln M-M-\sum_{k=0}N(k)[\ln N(k)-1+\ln k!]
\end{align}

\bigskip and the unbiased estimate is again given by maximizing $\Omega $
with respect to the distribution $N(k)$. The variational solution in this
case gives $-\ln N(k)-\ln k!+a-bk=0$, and thus $N(k)\sim \exp (-bk)/k!\sim
const^{k}/k!$. This is the Poisson distribution which in the context of
networks is associated with the random Erd\H{o}s-Renyi (ER) network.

As mentioned above a connected network corresponds to the case when all
boxes contain at least one ball. How does this constraint change the
distribution? It corresponds to first filling all the boxes with precisely
one ball. The number of ways of doing this is\ $M!/(M-N)!$. The number of
ways to fill in the remainder is given by 
\begin{equation}
\ln (M-N)!-\sum_{k=1}\ln N(k)!-\sum_{k=1}N(k)\ln f(k).
\end{equation}

Thus the crucial question is how the degeneracy $f(k)$ changes. Let
us start with ER-case and \textit{indistinguishable} balls. The degeneracy $%
f(k)$ is this time given by the number of distinct ways to fill a box with 
\textit{distinguishable} balls divided by the number of distinguishable ways
to fill a box will the remaining $k-1$ balls. For \textit{indistinguishable}
balls this is simple. There is only one distinguishable way to fill a box
with $k-1$ indistinguishable balls. Consequently, $f(k)=\frac{k!}{1}$ and
the distribution has again the Poisson-distribution form.

The case of \textit{distinguishable} balls is in this respect quite different: the
number of distinguishable ways to fill a box with $k-1$ \textit{%
distinguishable} balls is obviously $(k-1)!$ and consequently the degeneracy
is now $f(k)=\frac{k!}{(k-1)!}$. Thus the total number of distinguishable
ways of distributing \textit{distinguishable }balls subject to the
constraint that all boxes contains at least one ball is given by

\begin{align}
\ln \Omega =& \ln M!-\sum_{k}\ln N(k)!-\sum_{k}N(k)\ln k  \notag \\
\eqsim & M\ln M-M-\sum_{k}N(k)[\ln N(k)-1+\ln k].  \label{ways}
\end{align}%
The unbiased estimate of the total number of different states $\Omega $ is
given by the maximum of $\ln \Omega $ with respect to the distribution $%
N(k), $ subject to the constraints $\sum_{k=1}^{N}N(k)=N$ and $%
\sum_{k=1}^{N}kN(k)=M$ which by variational calculus gives $-\ln N(k)-\ln
k+a-bk=0$ with the solution 
\begin{equation}
N(k)=\frac{A\exp (-bk)}{k}
\end{equation}%
By this reasoning an unbiased random connected network has a
degree-distribution of the form

\begin{equation}
P(k)=\frac{N(k)}{N}\sim \frac{\exp (-bk)}{k}
\end{equation}

\textbf{To sum up}: We use the correspondence between the different ways to
distribute balls and states in statistical mechanics. Taking this into
account gives more possible states for distinguishable than for
indistinguishable balls. The link-ends on a node correspond to
distinguishable balls. An unbiased estimate assumes that all these states
are equally probable. This leads to a degree-distribution $P(k)\sim \exp
(-bk)/k$ which is distinctly different from the Poisson distribution.

How close is the analogy between distinguishable-balls-in-boxes (DBB) model
and a network? If we define a link by its link-ends such that an enumeration
of them are given by $(1,2),(3,4),...,(M-1,M)$ then the mapping is
precise. However, not all possible distributions of links fulfill the
requirement for a network. As pointed out above, the difference imposed by
the network constraints on the states of the DBB-model are often of minor
importance, except for the no-loop constraint which can be crucial. An
example when the no-loop condition makes a crucial difference between the
DBB-model and the corresponding network is the star-like networks (or more
generally networks which contain nodes of order $M$): the number of possible
states for such networks are significantly smaller in comparison with the
number of the corresponding DBB-states.

We note that a necessary \emph{but not sufficient} (because the orderings of
the objects in a box corresponds to different states) condition for two
states to be equal is that they contain precisely the same boxes provided a
box is identified by the specific balls it contains; for two equal
DBB-states all the boxes corresponding to the one state can be put on top of
identical boxes corresponding to the other. Likewise for two equal states of
a network: the two networks corresponding to the equal states can be put on
top of each other so that all nodes and all links precisely match.

How should one actually fill the boxes with balls in case of the constrained
DBB-model? One starts with one ball in each box. Next one randomly chooses
one of the remaining $M-N$ balls and add another ball in an arbitrary box.
This box is chosen with the probability $p(k=1)\sim 1$, $p(k>1)\sim k-1$ in
order to ensure that all the possible states are given the same chance. It
follows that

\begin{align*}
p(1) & =\frac{1}{N(1)+\sum(k-1)N(k)} \\
p(k>1) & =\frac{k-1}{N(1)+\sum(k-1)N(k)}
\end{align*}

Throwing the balls into the boxes according to the above probability
distribution gives the unbiased distribution $P(k)\sim\exp(-bk)/k$ (for $k>1$%
). If the balls are already in the boxes and one wants to choose all
possible states with equal probability, then one have to use the procedure
of choosing two balls randomly and then move the one to the same box as the
other \cite{seb06}. We also note in passing that for the unconstrained
DBB-model, the probability reduces to $p(k)\sim k$ which has the form of a
preferential attachment to boxes. However, in our context it arises as a
direct statistical consequence of distinguishable balls and in fact
represents the \emph{unbiased} situation with no preference what so ever!

\section{Box Information and Scale-freeness}

So far we have argued that the random distribution for a network is given by 
$P(k)\sim \exp (-bk)/k$. What type of bias in the sampling of different
states are necessary for changing the degree-distribution to a power law $%
P(k)\sim k^{\gamma }$ and what is the least bias necessary? To this end we
consider the box information (the information contained \emph{within} a box)
of the DBB-model. The box (or "useful") information, $s_{in}$, contained in
a box with distinguishable particles is given by the logarithm of the number
of possible different orderings of the balls within the boxes. For the
unconstrained DBB model this is $s_{in}=\ln k!$, while for the constrained
DBB-model it is $s_{in}=\ln (k-1)!$. The total box information for the
unconstrained DBB-model is hence $S_{in}=\sum_{k}N(k)\ln k!$. Note that this
is the maximum box information you can store for a given distribution $N(k)$%
. The global maximum is trivially the case when all balls are in the same
box \emph{i.e.}\ $N(M)=1$ and $N(k<M)=0$ so that $S_{in}^{\max }=\ln M!$.
For the constrained DBB-model we instead have $S_{in}=\sum_{k}N(k)\ln (k-1)!$
with the global maximum $S_{in}^{\max }=\ln (M-N+1)!$ for $N(M-N+1)=1$ and $%
N(1)=N-1$. The difference between the total box information for the
unconstrained DBB-model, \textit{i.e.} the maximum box information you can
store for a given distribution $N(k)$, and the total box information for the
constrained DBB-model (divided by the number of boxes to get an intensive
quantity) is 
\begin{equation}
\Delta S_{in}=\frac{1}{N}\sum_{k=1}N(k)\ln k.
\end{equation}

Thus small $\Delta S_{in}$ means large box information. It seems plausible
to us that a bias towards large box information could be favored in various
contexts and will hence consider how such a bias will affect the network
degree-distribution.

\begin{figure}[tbp]
\centering
\includegraphics[width=\columnwidth]{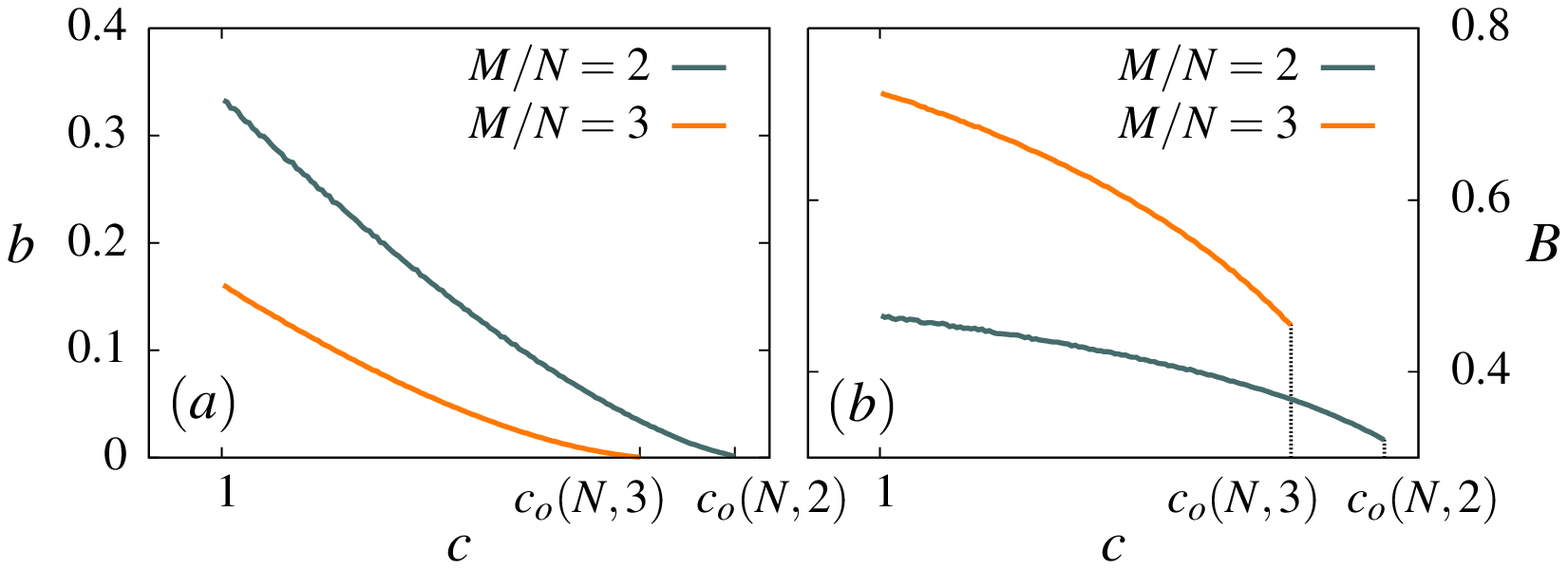} 
\caption{Figure showing how the Lagrange multiplier $b$ (Fig.\ a) and the
box information difference $B$ (Fig.\ b) depend on $c$ for two different
average number of balls ($2$ and $3$) and $N = 10^6$. Here, $c_0(N,2)=2.41$
and $c_0(N,3)=2.15$.}
\label{b_c}
\end{figure}

In order to find the distribution $N(k)$ which corresponds to the smallest $%
\Delta S_{in}$ we do a variational calculation: In addition to the basic
constraints for a non-growing network, $\sum_{k=1}^{N}P(k)=1$ and $%
\sum_{k=1}^{N}kN(k)=M/N$, we then also have the constraint $\Delta
S_{in}[P(k)]=B$. The fixed information value, $B$, introduces a bias on the
purely random states. This biased $P_{B}(k)$ is obtained by maximizing the
number of different states, subject to the three constraints. These
constraints are handled by three Lagrange multipliers and the solution is
the one which maximizes 
\begin{equation}
-\sum_{k}P_{B}(k)[\ln P_{B}(k)+a+bk+c\ln k]
\end{equation}%
with respect to the distribution $P_{B}(k)$. This leads to 
\begin{equation}
P_{B}(k)=\frac{A\exp (-bk)}{k^{c}}
\end{equation}%
where the constant $A$, $b$, and $c$ are given by the three constraints.
From the normalization condition $\sum_{k=1}^{N}P_{B}(k)=1$ we directly get
$A(b,c)$ and the remaining constants $b$ and $c$ are obtained from the
remaining two constraints. Figure \ref{b_c}(a) shows how $b$ and $c$ depend
on the ratio $M/N$. The possible solutions ranges from $(1\leq c<c_{0},\ b>0)
$ to $(c=c_{0},\ b=0)$ i.e.\ from $P(k)\sim \exp (-bk)/k^{c}$ to $P(k)\sim
k^{-c_{0}}$, where $c_{0}=c_{0}(N,M/N)$. [Note that $c_{0}>2$ for large N
because of the normalization condition]. Using the explicit form of the
solutions it is easy to show that for $M$ balls and $N$ boxes the scale-free
distribution corresponds to the smallest $B$ and hence to the largest box
information for the "least bias" solutions (see Fig.\ \ref{b_c}(b))\cite{bias}.

The relation between the global maximum solutions mentioned above (which are
not obtained within variational calculus) and networks will be discussed in
the following. Our conclusion will be that the scale-free network
corresponds to the global maximum of node information.

\section{Scale-freeness and Noise}

The degree distributions $P(k)$ are average distributions. This means that,
if $p_{i}$ is the probability for a possible distinct system state 
$i$, then the average distribution $P(k)$ is given by 
$P(k)=\sum_{i}p_{i}N_{i}(k)/N$ $=\langle n(k)\rangle$, where the sum is over all
box distributions corresponding to the system states $i$ (denoted by $%
\langle \rangle $). We have so far only characterized the system by the
average box distribution $\langle n(k)\rangle$. Our goal is to find as
detailed characteristics as possible for the maximum node-information random
scale-free network, in order to be able to decide whether or not a
particular network obeys the statistical properties implied by this
particular random scale-free network. To this end we also derive the
statistical deviations from the average distribution $\langle n(k)\rangle$.
This statistical deviation is measured by the noise 
\begin{equation}
\sum_{k}\langle (n(k)-\langle n(k)\rangle ^{2}\rangle .
\end{equation}

\bigskip The actual calculation follows the same steps as the variational
calculation of the maximum box information scale-free network in the
previous section. Again we address the problem of imposing the fixed
information constraint $\Delta S_{in}=B$ into the DBB-model 
\begin{equation}
B=\sum_{k}P(k)\ln k=\langle \sum n(k)\ln k\rangle
=\sum_{i}p_{i}\sum_{k}n_{i}(k)\ln k
\end{equation}%
Thus our constraints are now a fixed value $B,$ $\sum_{i}p_{i}=1$ and $%
\sum_{i}p_{i}\sum_{k}n_{i}(k)k=M/N.$ We want to maximize the average number
of distinct states $\Omega $ or equivalently 
\begin{equation}
\ln \Omega =const-\sum_{i}p_{i}\sum_{k}n_{i}(k)\ln [n_{i}(k)k]
\end{equation}%
(see Eq.\ (\ref{ways})). With no bias the problem just corresponds to
maximizing the entropy $-\sum_{i}p_{i}\ln p_{i}$ subject to the constraint $%
\sum_{i}p_{i}=1$ \cite{jaynes57}: the unbiased value of the probability for
a state is the maximum of $-\sum_{i}p_{i}[\ln p_{i}+const]$ with respect to
variations of $p_{i}$. The answer is obviously that all $p_{i}$ are equal
i.e.\ all system states are equally probable. Note that the variation is now
with respect to the probabilities $p_{i}$ of the system states $i$, whereas
in the previous sections it was with respect to the average distributions $%
P(k)$. For fixed values of $B$, $\Omega$ and $M/N$ the problem instead
corresponds to finding the maximum of the expression (where $\alpha$, $\beta$,
$\gamma$ and $\delta$ are Lagrange multipliers).

\begin{align}
& -\sum_{i}p_{i}[\ln p_{i}-\alpha \sum_{k}n_{i}(k)\ln k+\beta
\sum_{k}n_{i}(k)\ln [n_{i}(k)k]  \notag \\
& +\gamma \sum_{k}n_{i}(k)+\delta \sum_{k}n_{i}(k)k],
\end{align}%
with respect to variations of $p_{i}$. We here use different symbols for the
multipliers than in previous sections in order to emphazise that the
variations are with respect to a different variable. The result is
straightforwardly obtained and gives the probabilty for a system state $i$
as an exponential $p_{i}\sim \exp (-H/T)$ where

\begin{align}
\frac{H}{T}=&-\alpha \sum_{k}n_{i}(k)\ln(k)+\beta \sum_{k}n_{i}(k)\ln
[n_{i}(k)k]  \notag \\
&+\gamma \sum_{k}n_{i}(k)+\delta \sum_{k}n_{i}(k)k  \label{H1}
\end{align}

\bigskip Here we used the notation $\frac{H}{T}$ for the function in the
exponent in order to display the equivalence with statistical physics and
the Boltzman factor: in statistical physics the probability of a system
state $i$ is given by $p_{i}=Z^{-1}\exp (-H/T)$ where $H$ is the
hamiltonian, $T$ is the temperature and $Z^{-1}$ is a the normalization
constant determined by the condition $\sum_{i}p_{i}=1$. In statistical
physics $Z$ is called the partition function. The average value of any
quantity $O(n(k),k)$ is given by $\langle Z^{-1}\exp (-H/T)O(n(k),k)\rangle $
where the brackets means the average over all different states. For the
DBB-model the two last constraints in Eq.(\ref{H1}) (i.e. constant number of
boxes and balls) are included already in the definition of the model, so the
Hamiltonian $H$ reduces to

\begin{equation}
\frac{H}{T}=\frac{1}{T}[\sum_{k}n(k)\ln n(k)+\gamma_{0}\sum_{k}n(k)\ln k]
\label{H2}
\end{equation}

\begin{figure}[tbp]
\centering
\includegraphics[width=0.9\columnwidth]{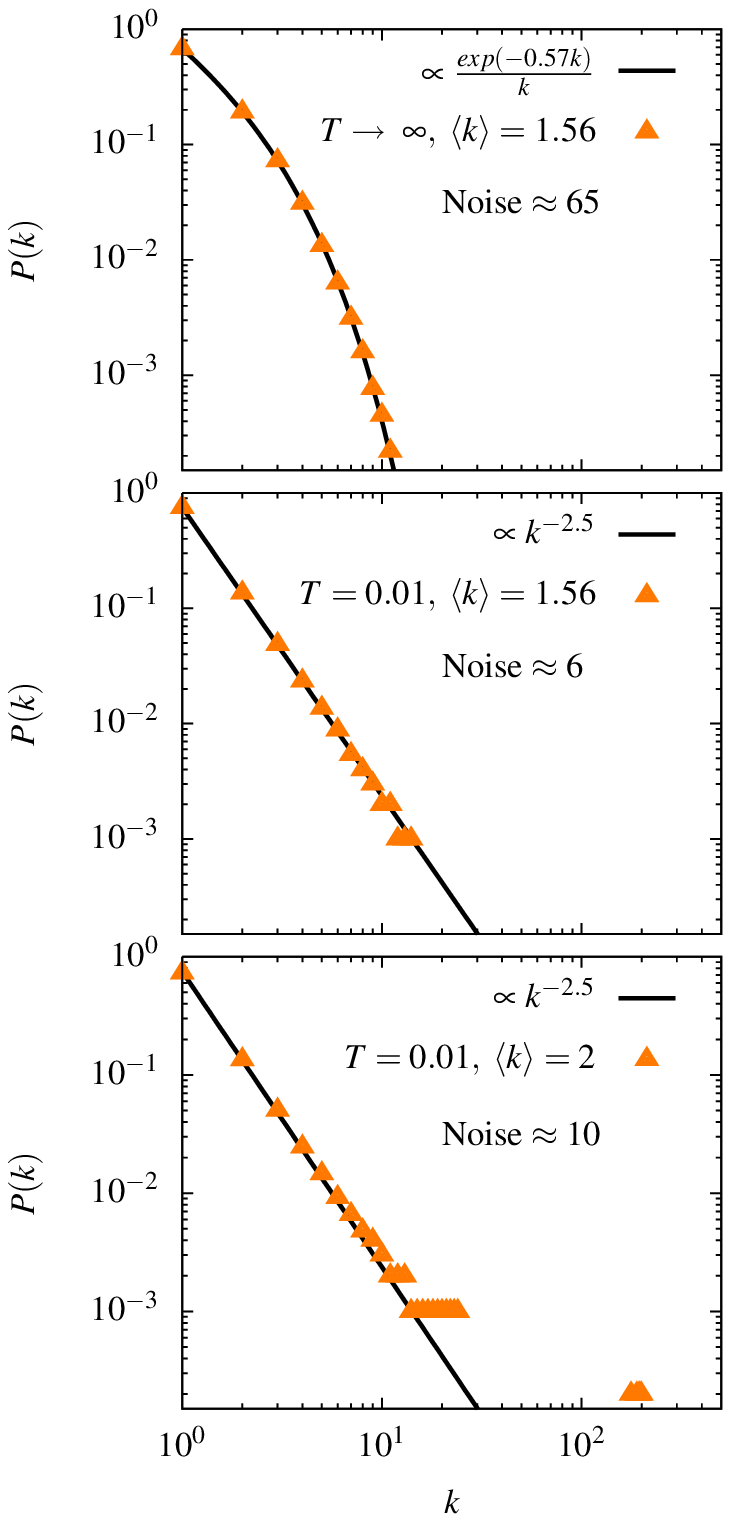}
\caption{Distribution of box-sizes obtained by Monte Carlo simulations where
the Hamiltonian from Eq.\ \ref{H2} was used. $N = 1000$, $\protect%
\gamma_0=2.5$ and the results were averaged over $2000$ runs. Figure (a) and
(b) shows the results for an average number of balls equal to $1.56$ and $T$
reaching infinity and zero respectively. In (c), $T$ is also close to zero
but $M/N = 2$ and boxes (nodes) of order $M$ is forming. Note that the
Scale-free distribution (b) has the smallest noise.}
\label{mc}
\end{figure}

with $T=1/\beta $ and $\gamma _{0}=(\beta -\alpha )/\beta $. This system
state probability and Hamiltonian in the network context was first derived
and investigated in Ref.\ \cite{seb06}. In particular Monte-Carlo
simulations are easily implemented following the standard proceedure in
statistical physics: choose two random balls and then move one to the same
box as the other with a probability given by the Boltzmann factor $\exp
(-H/T)$. Figure \ref{mc}(a) shows the unbiased estimate obtained from MC
simulations which obviously corresponds to $T=\infty $ and also $\alpha =0$
(= no bias). This is just the unconstrained DBB-model and the average
distribution is of the form $P(k)\sim \exp (-\delta k)/k$ as previously
shown. The new characteristics is the statistical deviations from the
average distribution, the noise, associated with the least bias (which in
this case is no bias at all). The least bias always corresponds to the
smallest ratio $\alpha /\beta $ in Eq.\ (\ref{H1}) and hence to the largest $%
\gamma _{0}$ in Eq.\ (\ref{H2}) for which the average solution of the form $%
P(k)\sim \exp (-\delta k)/k^{\gamma }$ with $1\leq \gamma <\gamma _{0}$ is
obtained. It turns out that for each given ratio $M/N$ there is precisely
one such largest value $\gamma _{0}$ which is given by the condition 
\begin{equation}
C\equiv \frac{\sum_{k=1}^{N}kk^{-\gamma _{0}}}{\sum_{k=1}^{N}k^{-\gamma _{0}}%
}=M/N
\end{equation}%
(note that $\gamma _{0}=c_{0}(N,M)$ where $c_{0}(N,M)$ is defined in the
preceding section). This also means that the statistical noise for the whole
sequence of least biased solutions is \textit{uniquely} given. Since $\gamma
_{0}$ is unique the noise is just given by the "temperature" $T$ : large
noise corresponds to large $T$ and small noise to small, which is, of
course, precisely what you would expect from statistical physics. The
scale-free distribution is obtained in the limit of small $T$ and
consequently has the smallest noise. Figure \ref{mc}(b) shows the average
distribution and the noise obtained for a low $T$. The important point is
the connection between the scale-free distribution and the smallest noise.

What do the two cases $C>M/N$ and $C<M/N$ correspond to? In the first case
the average distribution is of the form $P(k)\sim\exp(-\delta k)/k^{\gamma}$
with $\delta>0$. So the solutions overlap with the ones obtained for the
least bias. The point is that these solutions are obtained for a smaller $%
\gamma_{0}$ and hence for a larger bias. This means the same $P(k)$ but a
smaller noise. The case $C<M/N$ means that $\gamma_{0}$ is larger than for
the least bias solutions. In this case boxes with of order $M$ balls are
created. An example is given in Fig.\ \ref{mc}(c). Note that these solutions
with order $M$ nodes are not picked up by variational calculus.

For the network, the presence of nodes of order $M$ greatly constrains the
number of possible different states relative to the DBB-model, because of
the no-loop constraint. For example a perfect star with $N$ nodes and $%
M=2(N-1)$ link-ends [ $N-1$ single nodes attached to a central node of
degree $N-1)$] for the constrained DBB model corresponds to the box
information 
\begin{equation}
\Delta S_{in}=\sum_{k=1}P(k)\ln k\approx\frac{1}{N}\ln(N-1)\rightarrow0
\end{equation}
for large $N$. However, only one of these states is consistent with the
network requirement so that 
\begin{equation}
\Delta S_{in}\sim\frac{1}{N}[M\ln M-N\ln N]\approx\ln(N).
\end{equation}

For a network this implies the solutions containing order $M$ nodes will
always have much smaller node information than the corresponding scale-free
network.

Based on these observations, we conjecture that the scale-free network
maximizes the node information.

\section{Consequences for Directed Networks.}

So far we discussed undirected networks. However, also directed networks has
a one-to-one mapping to the constrained DBB-model. In this case a link is
again defined by two balls: if the balls are enumerated $1,2,,,M$ then the
links are enumerated by two consecutive numbers $12,34,,,\ M-1M$ where the
first number (which is always odd) denotes the start of the link and the
last (even numbers) denote the end of the link. This does not change
anything as to the number of different ways you can distribute the balls (or
links) among the boxes (nodes). Thus the scale-free network should again be
the one which maximizes the node information. Note that the direction of the
links does not affect the amount of node information which a node can carry.
Consequently, from the point of view of node information the directions of
the links attached to a node are completely random. Thus a network, which is
only optimized with respect to the node information, acquires characteristic
random features with respect to the distribution of \emph{in-going} and 
\emph{out-going} links attached to a node.

Some of these characteristic random features for \emph{in-} and \emph{out}%
-links attached to a node are as follows: Suppose that the numbers of the 
\emph{in}-, \emph{out}-, and \emph{total} numbers of links connected to a
node are $k_{in},$ $k_{out}$, and $k$ , respectively, so that $%
k_{in}+k_{out}=k$. The average number of \emph{in}-links $\langle
k_{in}\rangle_{out}$ for for nodes with \emph{precisely the number of }$%
k_{out}$ \emph{out}-links are then given by the binomial coefficient $%
B(k_{in},k)$ i.e.\ the probability to get $k_{in}$ tails when tossing a coin 
$k=k_{\mathrm{in}}+k_{\mathrm{out}}$ times:

\begin{equation}
\langle k_{in}\rangle_{k_{\mathrm{out}}}=\frac{%
\sum_{k=k_{out}}^{k_{max}-k_{out}}B(k-k_{out},k)P(k)(k-k_{out})}{%
\sum_{k=k_{out}}^{k_{max}-k_{out}}B(k-k_{out},k)P(k)}.  \label{kin}
\end{equation}

For a the case of $P(k)\propto1/k^{\gamma}$ one then finds $\langle k_{%
\mathrm{in}}\rangle_{out}\approx k_{\mathrm{out}}$ \cite{bernhardsson06}.
The result $\langle k_{\mathrm{in}}\rangle_{out}$ $\approx$\ $k_{\mathrm{out}%
}$ may look innocent, but it is completely non-trivial, as realized when
comparing to the ER-network: in this latter case there is no correlation and 
$\langle k_{\mathrm{in}}\rangle_{out}=const$ no matter what size $k_{out}$
one chooses. Another characteristic feature is the spread of the
distribution of $k_{in}$-links for the nodes with a given number of $k_{out}$%
. For this spread we use the measure 
\begin{equation}
S_{in}(k_{out})=\frac{\sum_{(k_{in}|k_{out})}|k_{in}-\langle
k_{in}\rangle_{k_{out}}|}{\langle k_{in}\rangle_{k_{out}}}
\end{equation}

which in terms of the binomial coefficient becomes%
\begin{align}
& S_{in}(k_{out})=  \notag \\
& \frac{\sum_{k=k_{out}}^{k_{max}-k_{out}}B(k-k_{out},k)P(k)\frac {%
k-k_{out}-\langle k_{in}\rangle_{k_{out}}}{\langle k_{in}\rangle_{k_{out}}}}{%
\sum_{k=k_{out}}^{k_{max}-k_{out}}B(k-k_{\mathrm{out}},k)P(k)}  \label{binom}
\end{align}

This gives $S_{\mathrm{in}}(k_{\mathrm{out}})\propto k_{\mathrm{out}}^{-1/2}$%
,whereas for the ER-network the spread is independent of $k_{\mathrm{out}}$
i.e.\ $S_{\mathrm{in}}(k_{\mathrm{out}})=const$ \cite{bernhardsson06}. From
symmetry one has $P_{in}(k)=P_{out}(k).$ Using the relation $\langle k_{%
\mathrm{in}}\rangle _{out}$ $\approx $\ $k_{\mathrm{out}}$ one can motivate
the (at least) approximate relation 
\begin{equation}
P_{tot}(k)\sim P_{in}(\frac{k}{2})=P_{out}(\frac{k}{2})
\end{equation}%
for even $k$, where $P_{tot}$, $P_{in}$ and $P_{out}$, are the size
distributions of the \emph{tot}-, \emph{in}-, and \emph{out}-links,
respectively. This means that all three distributions are described by the
same functional form $P(k)=P_{tot}(k)\sim P_{in}(\frac{k}{2})=P_{out}(\frac{k%
}{2}).$ The motivation steps are as follows:

\begin{align}
P_{in}(k_{in}) & = \sum_{k_{out}}P_{tot}(k_{out}+k_{in})\sim P_{tot}(\langle k_{out}\rangle |_{k_{in}}+k_{in}) \nonumber \\
& \simeq P_{tot}(2k_{in})
\end{align}

Taking into account that $\sum_{k}P_{tot}(2k)$ only has half as many points
as $\sum _{k}P_{in}(k)$ fixes the normalization constant to $2$ and thus
leads to the relation

\begin{equation}
2P_{tot}(2k)=P_{in}(k)
\end{equation}

The general connection between the \emph{total}- and \emph{in}-, \emph{out-}%
link distributions, 
\begin{equation}
P_{tot}(k_{tot})={\textstyle\sum\limits_{k_{in+k_{out}}=k_{tot}}}%
P_{in}(k_{in})P(k_{out}),
\end{equation}
is then of the form 
\begin{equation}
2\ P(2k)= {\textstyle\sum\limits_{q=0}^{k}} P(q)P(k-q)\text{ \ \ \ }for\text{
\ }k\geqq2.
\end{equation}

\subparagraph{Metabolic networks:}

\begin{figure}[tbp]
\centering
\includegraphics[width=\columnwidth]{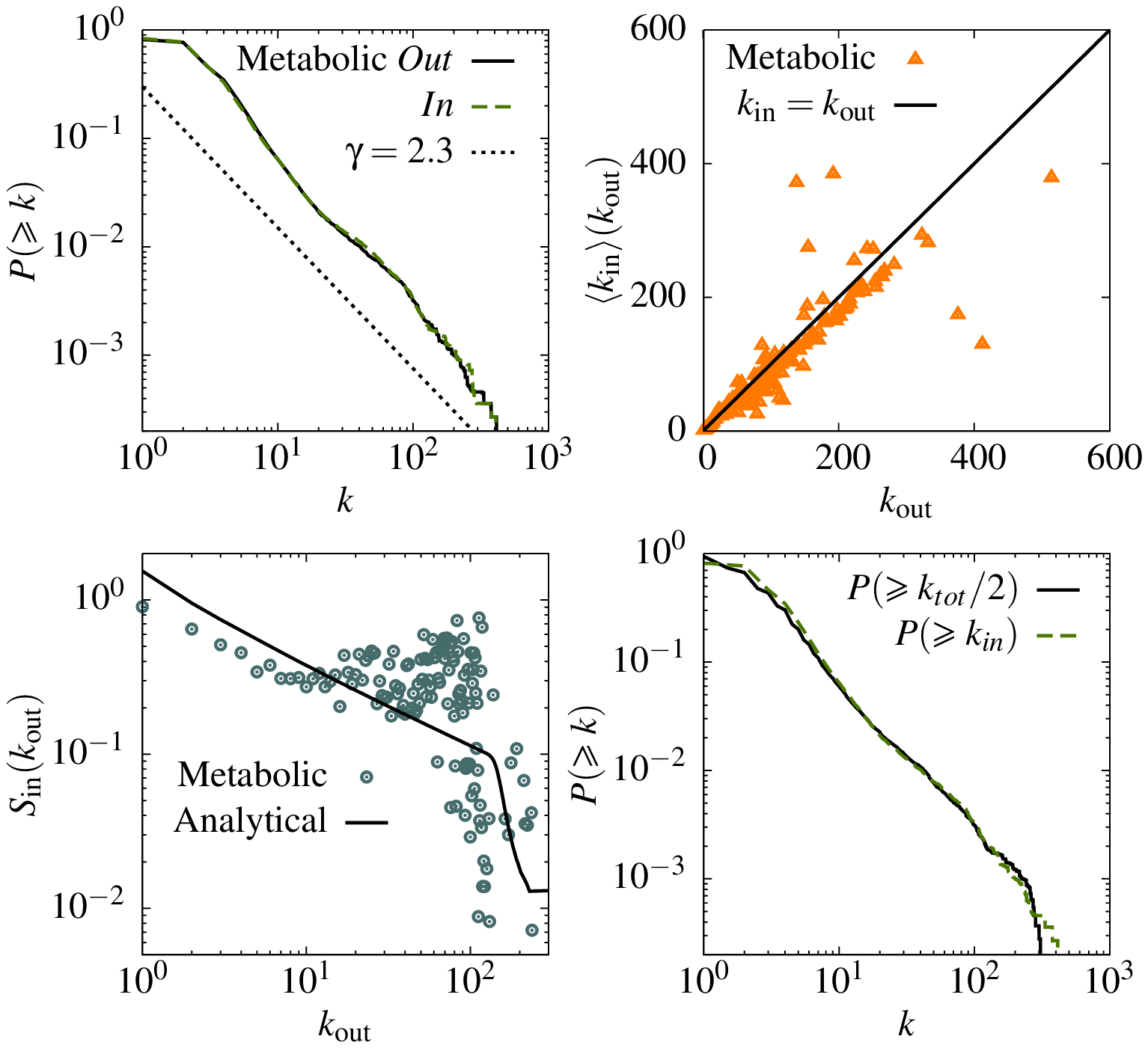} 
\caption{Metabolic Networks: Average over 107 metabolic networks with data
obtained from Ref.\ (\protect\cite{ma03a}). The same characteristics as for
Model A and Friendly Merging: a) Cumulative degree-distribution $P(\geq k)$.
The dashed straight line line has the slope $\protect\gamma=2.3$ and $%
\langle k_{in}\rangle=\langle k_{out}\rangle \approx4.3$. b) Plot of $%
\langle k_{in}\rangle_{out}$ versus $k_{out}$ showing that the data is
consistent with $\langle k_{in}\rangle_{out}=k_{out}$. c) Demonstration that
the spread goes as $S_{in}(k_{out})\propto k_{out}^{-1/2}$ to reasonable
approximation, full line from Eq.\ (\protect\ref{binom}) and data points
from simulations. The cut-off for large $k_{out}$ is a finite size effect.
d) demonstrates that the relation $2P_{tot}(2k)=P_{in}(k)=P_{out}(k)$ is
fulfilled as expected for a maximum-node-information network (note that $%
P(>k)=\sum_{k}^{k_{\max}}P(k)$).}
\label{metab}
\end{figure}

These three characteristics can now be used to compare with real networks.
As an illustration we choose metabolic networks:\cite{bernhardsson06} The
data shown in Fig.\ \ref{metab} are average properties of 107 metabolic
networks with the average size $\langle N\rangle\approx940$ (data taken from
Ref.\ \cite{ma03a}\cite{ma03b}). A metabolic network is constructed as
follows: Substrates and products in the metabolism are nodes. Two nodes are
connected if the substance of one is a substrate in a metabolic reaction
which produces the substance represented by the other node. The links points
from the substrate to the product. The data are obtained as the average over
107 such networks and consequently reflect an ensemble average network
structure associated with metabolic networks. Figure \ref{metab}(a) shows
that metabolic networks have $P_{out}(k_{out})=P_{in}(k_{in})$ and also have
a broad scale-free degree-distribution as was first demonstrated in Ref.\ 
\cite{jeong00}. In addition from Fig.\ \ref{metab}(b) and (c) we conclude
for the ensemble average of metabolic networks the relation $\langle
k_{in}\rangle\approx k_{out}$ holds to good approximation in accordance with
the maximum node-information requirement. Likewise the spread has a similar
decrease as required by maximum node information. Furthermore, Fig.\ \ref%
{metab}(d) shows that the relation $2P_{tot}(2k)=P_{in}(k)=P_{out}(k)$ holds
to excellent approximation [in accordance with standard practice in network
analysis contexts, it is in Fig.\ \ref{metab}(d) expressed in terms of the
cumulant distribution $P(>k)=\sum_{k}^{k_{\max}}P(k)$]. Thus ensemble
averages over metabolic networks show every sign of belonging to a maximum
node-information network.

\section{Concluding Remarks}

In this paper we introduced and studied the constrained DDB-model. This
model has a one-to-one mapping to a network provided the network is allowed
to have loop-links and to be disconnected. We showed that the maximum box
information state for the constrained DBB-model is a scale-free
size-distribution of the boxes described by a power-law with an exponent $%
-\gamma$ with $\gamma>2$, \emph{provided} the possibility of having boxes
with a finite fraction of the balls are excluded (exclusion of order $M$
boxes). Next we observed that the presence of order $M$ nodes is effectively
suppressed for connected networks with no loop-links. This led to the
conjecture that the scale-free network gives the maximum node information.
This conjecture in turn leads to explicit characteristic features for the
scale-free maximum node-information network. We showed that an ensemble of
metabolic networks has these features and concluded that metabolic network
appears to have evolved in a way which maximizes the node information. The
obvious question is then "Why?". We have at present no answer to this apart
from the general observation\ that keeping as many options as possible
increases the chances to adapt to new conditions. Neither has it escaped our
notice that in the case of metabolic networks the ordering of the link ends
on a node corresponds to a time-ordering and that further a particular
time-ordering is required for creating a specific functional metabolic
pathway. If this is the case one might speculate that maximum node
information corresponds to the maximum number of potential metabolic
pathways.

\begin{acknowledgement}
This work was supported by the Swedish research Council through contract
50412501.
\end{acknowledgement}


\begin{thebibliography}
\fi
\expandafter\ifx\csname bibnamefont\endcsname\relax

\fi
\expandafter\ifx\csname bibfnamefont\endcsname\relax

\fi
\expandafter\ifx\csname citenamefont\endcsname\relax

\fi
\expandafter\ifx\csname url\endcsname\relax

\fi
\expandafter\ifx\csname urlprefix\endcsname\relax

\fi
\providecommand{\bibinfo}[2]{#2} \providecommand{\eprint}[2][]{\url{#2}}

\bibitem[Albert and Barab\'{a}si(2002)]{albert02} \bibinfo{author}{%
\bibfnamefont{R.}~\bibnamefont{Albert}} and \bibinfo{author}{%
\bibfnamefont{A.-L.} \bibnamefont{Barab\'{a}si}}, \bibinfo{journal}{%
\mbox{Rev. of Mod. Phys.}} \textbf{\bibinfo{volume}{74}}, \bibinfo{pages}{47}
(\bibinfo{year}{2002})

\bibitem[Dorogovtsev and Mendes(2003)]{dorog03} \bibinfo{author}{%
\bibfnamefont{S.}~\bibnamefont{Dorogovtsev}} and \bibinfo{author}{%
\bibfnamefont{J.}~\bibnamefont{Mendes}}, \emph{%
\bibinfo{title}{Evolution of Networks:From Biological Nets to the
  Internet and WWW}} (\bibinfo{publisher}{Oxford University Press}, %
\bibinfo{year}{2003})

\bibitem[Newman(2003)]{newman03} \bibinfo{author}{\bibfnamefont{M.E.J}~%
\bibnamefont{Newman}}, \bibinfo{journal}{\mbox{SIAM Review}} \textbf{%
\bibinfo{volume}{45}}, \bibinfo{pages}{167} (\bibinfo{year}{2003}).

\bibitem[Boccaletti et~al.(2006)]{bocca06} \bibinfo{author}
{\bibfnamefont{S.}~\bibnamefont{Boccaletti}}, %
\bibinfo{author}{\bibfnamefont{V.}~\bibnamefont{Latora}}, %
\bibinfo{author}{\bibfnamefont{Y.}~\bibnamefont{Morena}}, %
\bibinfo{author}{\bibfnamefont{M.}~\bibnamefont{Chavez}}, and %
\bibinfo{author}{\bibfnamefont{D.-U.} \bibnamefont{Hwang}}, %
\bibinfo{journal}{\mbox{Phys. Rep.}} \textbf{\bibinfo{volume}{424}}, %
\bibinfo{pages}{175} (\bibinfo{year}{2006}).

\bibitem[Barab\'{a}si, Albert and Jeong(1999)]{barabasi99} %
\bibinfo{author}{\bibfnamefont{A.-L.} \bibnamefont{Barab\'{a}si}}, %
\bibinfo{author}{\bibfnamefont{R.}~\bibnamefont{Albert}}, and %
\bibinfo{author}{\bibfnamefont{H.}~\bibnamefont{Jeong}}, \bibinfo{journal}{%
\mbox{Science}} \textbf{\bibinfo{volume}{286}}, \bibinfo{pages}{509} (%
\bibinfo{year}{1999}).

\bibitem[Valverde et~al.(2002)]{valverde02} \bibinfo{author}{%
\bibfnamefont{S.}~\bibnamefont{Valverde}}, \bibinfo{author}{%
\bibfnamefont{R.F.}~\bibnamefont{Cancho}} and \bibinfo{author}{%
\bibfnamefont{R.V.}~\bibnamefont{Sol\'{e}}}, \bibinfo{journal}{%
\mbox{Europhys. Lett}} \textbf{\bibinfo{volume}{60}}, \bibinfo{pages}{512} (%
\bibinfo{year}{2002}).

\bibitem[Kim et~al.(2005)]{beom05} \bibinfo{author}{\bibfnamefont{B.J.}~%
\bibnamefont{Kim}}, \bibinfo{author}{\bibfnamefont{A.}~\bibnamefont{Trusina}}%
, \bibinfo{author}{\bibfnamefont{P.}~\bibnamefont{Minnhagen}}, and %
\bibinfo{author}{\bibfnamefont{K.}~\bibnamefont{Sneppen}}, %
\bibinfo{journal}{\mbox{Eur. Phys. J. B}} \textbf{\bibinfo{volume}{43}}, %
\bibinfo{pages}{369} (\bibinfo{year}{2005}).

\bibitem[Thurner and Tsallis(2005)]{thurner05} \bibinfo{author}{%
\bibfnamefont{S.}~\bibnamefont{Thurner}} and \bibinfo{author}{%
\bibfnamefont{C.}~\bibnamefont{Tsallis}}, 
\bibinfo{journal}{\mbox{Europhys.
Lett}} \textbf{\bibinfo{volume}{72}}, \bibinfo{pages}{197} (%
\bibinfo{year}{2005}).

\bibitem[Bianconi(2006)]{Bianconi} \bibinfo{author}{\bibfnamefont{G. Bianconi}},
\bibinfo{journal}{\mbox{cond-mat/0606365v5}} (\bibinfo{year}{2006})

\bibitem[Jaynes(1957)]{jaynes57} \bibinfo{author}{\bibfnamefont{E.T.}~%
\bibnamefont{Jaynes}}, \bibinfo{journal}{\mbox{Phys. Rev}} \textbf{%
\bibinfo{volume}{106}}, \bibinfo{pages}{620} (\bibinfo{year}{1957})

\bibitem[Bialas et~al.(2000)]{bialas00} 
\bibinfo{author}{\bibfnamefont{P.}
\bibnamefont{Bialas}}, \bibinfo{author}{\bibfnamefont{L.}~%
\bibnamefont{Bogacz}}, \bibinfo{author}{\bibfnamefont{Z.}~%
\bibnamefont{Burda}}, and \bibinfo{author}{\bibfnamefont{A.}~%
\bibnamefont{Krzywicki}}, \bibinfo{journal}{\mbox{Nucl. Phys. B}} \textbf{%
\bibinfo{volume}{575}}, \bibinfo{pages}{599} (\bibinfo{year}{2000}).

\bibitem[Burdas et~al.(2001)Burdas, Correia, and Krzywicki]{burda01} %
\bibinfo{author}{\bibfnamefont{Z.}~\bibnamefont{Burdas}}, %
\bibinfo{author}{\bibfnamefont{J.}~\bibnamefont{Correia}}, and %
\bibinfo{author}{\bibfnamefont{A.}~\bibnamefont{Krzywicki}}, %
\bibinfo{journal}{\mbox{Phys. Rev. E}} \textbf{\bibinfo{volume}{64}}, %
\bibinfo{pages}{046118} (\bibinfo{year}{2001}).

\bibitem[Minnhagen et~al.(2005)]{seb06} \bibinfo{author}{\bibfnamefont{P.}~%
\bibnamefont{Minnhagen}}, \bibinfo{author}{\bibfnamefont{S.}~%
\bibnamefont{Bernhardsson}}, and \bibinfo{author}{\bibfnamefont{B.J.}~%
\bibnamefont{Kim}}, \bibinfo{journal}{\mbox{Submitted to Eur. Phys. Lett.}} (%
\bibinfo{year}{2006}).

\bibitem[Bernhardsson and Minnhagen(2006)]{bernhardsson06} %
\bibinfo{author}{\bibfnamefont{S.}~\bibnamefont{Bernhardsson}} and %
\bibinfo{author}{\bibfnamefont{P.}~\bibnamefont{Minnhagen}} %
\bibinfo{journal}{\mbox{Phys. Rev. E}} \textbf{\bibinfo{volume}{74}}, %
\bibinfo{pages}{026104} (\bibinfo{year}{2006}).

\bibitem[Ma et~al.(2003)]{ma03a} \bibinfo{author}{\bibfnamefont{H.}~%
\bibnamefont{Ma}}, and \bibinfo{author}{\bibfnamefont{A.-P.}~%
\bibnamefont{Zeng}}, \bibinfo{journal}{\mbox{Bioinformatics}} \textbf{%
\bibinfo{volume}{19}}, \bibinfo{pages}{270} (\bibinfo{year}{2003}).

\bibitem[Ma et~al.(2003)]{ma03b} \bibinfo{author}{\bibfnamefont{H.}~%
\bibnamefont{Ma}}, and \bibinfo{author}{\bibfnamefont{A.-P.}~%
\bibnamefont{Zeng}}, \bibinfo{journal}{\mbox{Bioinformatics}} \textbf{%
\bibinfo{volume}{19}}, \bibinfo{pages}{1423} (\bibinfo{year}{2003}).

\bibitem[Jeong et~al.(2000)]{jeong00} \bibinfo{author}{\bibfnamefont{H.}~%
\bibnamefont{Jeong}}, \bibinfo{author}{\bibfnamefont{B.}~%
\bibnamefont{Tombor}}, \bibinfo{author}{\bibfnamefont{R.}~%
\bibnamefont{Albert}}, \bibinfo{author}{\bibfnamefont{Z.N.}~%
\bibnamefont{Oltvai}}, and \bibinfo{author}{\bibfnamefont{A.-L}~%
\bibnamefont{Barab\'{a}si}}, \bibinfo{journal}{\mbox{Nature}} \textbf{%
\bibinfo{volume}{407}}, \bibinfo{pages}{651} (\bibinfo{year}{2000}).

\bibitem[Least bias(2006)]{bias} \bibinfo{}{The ``least bias'' solution 
corresponds to the largest number of states for a given value of the
entropy $-\sum_k P(k)ln P(k)$ see Ref.\ [13].}

\end{thebibliography}

\end{document}